# Development of a Blockchain-Based On-Demand Lightweight Commodity Delivery System

Bayezid Al Hossain Onee, Kaniz Fatema Antora, Omar Sharif Rajme, and Nafees Mansoor

University of Liberal Arts Bangladesh (ULAB), 688 Beribadh Road, Dhaka 1207, Bangladesh

bayejid.hossain.cse@ulab.edu.bd,
kaniz.fatema.cse@ulab.edu.bd, omar.sharif.cse@ulab.edu.bd,
nafees.mansoor@ulab.edu.bd

\* Corresponding Author: Nafees Mansoor

**Abstract.** The COVID-19 pandemic has caused a surge in the use of online delivery services, which rely on user-generated content to promote collaborative consumption. Although Online Food Delivery (OFD) is a popular delivery system in Bangladesh, it has yet to ensure item authenticity, especially with the increasing demand for lightweight commodity delivery services across the country. The authenticity of products, involvement of multiple parties, and fair exchange are all challenging aspects of coast-to-coast services. Therefore, it is necessary for the three entities involved in the supply chain transaction - Seller, Carrier, and Buyer - to establish at least two peer-to-peer operations to ensure reliability and efficiency. To address these limitations and meet consumer expectations, the study proposes a framework for a nationwide on-demand marketplace for lightweight commodity items and a delivery system. Furthermore, transaction details are stored in a Blockchain to ensure the transparency and reliability of the proposed system.

## 1   Introduction

The development of technology is bringing about significant changes across various fields, including marketplaces. However, while Software Change Management (SCM) plays a critical role in an organization's overall performance, it has not kept pace with technological advancements. Consequently, the current supply chain faces challenges in meeting people's demands for product quality, authenticity, fair pricing, and relevant information. Moreover, companies with extended global supply chains incur between 80 to 90 percent of their costs in supply chain management [1], which can lead to product price hikes and a compromise on product quality [2]. The involvement of more intermediaries in SCM increases system complexity and results in longer delivery durations and higher costs. Each intermediary in the supply chain seeks to profit, leading to the injection of adulteration into raw materials for additional profit without the manufacturer's permission [3]. To address these issues, this study proposes an on-demand lightweight commodity item marketplace and a delivery system.

The traditional delivery system has several limitations, such as being costly and time-consuming, limited accessibility in rural areas, lack of proper tracking, and environmental impact due to carbon emissions. Additionally, the system may need to be more flexible to accommodate last-minute changes or custom customer requests. It may offer limited delivery options, such as same-day or weekend delivery. Lastly, the traditional delivery system may need proof of the goods' condition during transit.

The proposed lightweight commodity delivery system is cost-effective for less fuel, maintenance, and operational costs than traditional delivery systems. These systems are quicker to transport goods, more flexible to quickly adopt changes in delivery locations, more convenient for offering more flexible delivery options, and can deliver packages directly to a customer's doorstep. Most importantly, the lightweight commodity delivery system provides more efficiency, lower emissions, and less environmental impact. Crowd-sourcing-based carrier hiring is also being considered to ensure the smooth execution of the platform. This involves hiring registered users as carriers to deliver products from the origin of the requested product to the consumer's destination. This not only ensures timely delivery but also creates job opportunities for individuals traveling across the country, especially in developing countries where unemployment rates are high.

The proposed system guarantees the privacy of user data through a decentralized database, numerous security protocols, and algorithms that authenticate and maintain the unmodified nature of user data. Moreover, the user-friendly design of the platform makes it easily accessible through both smartphone apps and websites, ensuring a hassle-free experience for customers. Additionally, the proposed system ensures a fair exchange of products and enhances the stability of the supply chain. By ensuring the authenticity of products and transparency in the delivery process, the platform can boost the confidence of customers, thereby increasing demand for authentic products. This, in turn, can promote the economic development of developing countries and ensure the stable, authentic production of products in those countries. The objective of this study is to provide

- A blockchain-based online lightweight commodity delivery system where customers can order fresh foods.
- A cost-effective business model with a higher security protocol that works on both android and web portals.
- Provide an online delivery system without third-party interference.

The rest of this article is structured as follows: Section 2 provides a comprehensive examination of the relevant literature pertaining to the existing on-demand delivery systems. In Section 3, the architecture of the proposed system is discussed. Next, the proposed system is explained in Section 4, whereas the design and development of the proposed system are presented in Section 5. finally, the article is concluded in Section 6 along with the future works.

## 2 Existing Systems

In Bangladesh as well as in other countries, there exist quite a few on-demand delivery systems. Some of the popular systems are discussed in this section highlighting both their strengths and limitations.

Foodpanda is the most familiar food delivery system across 50 countries worldwide, headquartered in Berlin, Germany [4]. Its main job is to enable users to place orders at nearby restaurants with the help of its website or mobile app. There are some existing delivery services if we consider the delivery system of solid products such as electronic gadgets, resources of complete products, crafting products, etc. They allow the consumer to deliver products across the country, such as Sundarban Courier Service, S.A. Paribahan LTD, etc. Their main target is to deliver solid products or any digestive product. Although they deliver those types of products, their service could be better, which is not usable for daily purposes.

Pathao Food and Shohoz Food are two Bangladeshi companies offering food delivery services, with Pathao launching its service in January 2018 in Dhaka and Chittagong, and Shohoz Food starting in October 2018 [5]. Pathao Food is a Bangladeshi food delivery system that operates similarly to Foodpanda but does not have a restaurant agreement. Shohoz Food is another Bangladeshi food delivery system that began operations in October 2018 [6]. ShipChain is a logistic utility ecosystem that leverages blockchain technology to provide an integrated system for logistics companies, allowing for benefits such as trustless contract execution, historical data immutability, and no single point of failure [7]. Triwer is a Norwegian company aiming to eliminate inefficiencies in cargo delivery by introducing a crowd delivery marketplace and recording intermediate transactions, parcel tracking, and pricing on their custom side-chain while running the delivery smart contracts on the Ethereum blockchain. Triwer plans to charge a commission of 5-15 percent on all delivery revenue [8].

The Israeli PAKET Project aims to establish a decentralized delivery network that is accessible to all, transparent, and does not charge any commission. The project is developing a protocol that enables physical parcel deliveries among network participants, regardless of their location. The PAKET protocol ensures the safe delivery of parcels through collateral deposits, relays, and storage in hubs [8]. The project partners offer a Blockchain-secured parcel service model that complements the logistics system. This model ensures reliable tracking of package delivery using a Smart Contract triggered by an off-chain event. The delivery agent serves as a reliable "oracle" to confirm delivery to the buyer or self-service package station, and a multi-signature procedure adds an extra layer of fraud protection. The parcel service also records the sender's ID and ensures that personal data is not stored on the Blockchain but is cryptographically secured. In case of privacy constraints, the seller's identity can be entirely hidden from the logistics operator [9].

The company, ShopUp, operates in Bangladesh and facilitates connections between producers and wholesalers. ShopUp addresses issues such as product unavailability, lack of transparent pricing, and inefficient delivery systems that hinder small entrepreneurs in their daily business operations. The company has developed a B2B platform to enable rapid connections between producers, wholesalers, retailers, and consumers. ShopUp has successfully migrated a significant portion of the market to digital platforms and additionally provides services such as last-mile logistics, digital credit, and business management solutions [10].

All of these systems are involved in the delivery system. However, each of their working principles differs in some sectors, but mainly they work in the same way. Some serve in the local area, such as Pathao Food, Shohoz Food, and FoodPanda. They can only deliver products in a particular city, which creates a problem of product availability across the country. Other issuers take charge of registering in their system, such as food panda, which charges $100-$150 per restaurant as a registration fee. Then Foodpanda charges a 15-20 percent commission on every food order the restaurant receives [11]. Also, there is a delivery charge depending on the distance.

According to some research, the total daily delivery was at 25000 orders per day in 2019 on average [12]. There is an estimate that the overall market size for food delivery is $10 million, and it could grow to over $5 billion by 2025 [13]. Also, another major issue with delivery services, such as Sundarban courier services which serve across the country delivery system, needs a real-time tracking service. The consumer needs help to track their product in real time. Consumers need to depend on their agent or ask their customer service center every day about their product. They need to find out where the product comes from, the manufacturing element of their buying product, who maintains the product cycle, etc. After analyzing the problems, we found a big issue: each product needs an authorized document from the particular producer of each product. So that a consumer buys a product that depends on only vendor guarantee, which is valueless. Moreover, the product is terrible for the consumers and can be adulterated or cannot be long-lasting because of artificial manufacturing products.

## 3   System Architecture

The conceptual framework and architecture of the proposed system are presented herein:

This section elaborates on the design elements of the developed product delivery system and the specific intentions behind its design (Fig. 4). The system comprises various technologies, including a private blockchain server, a decentralized database for authenticated users, a mobile and web application for supply chain management, and a monitoring system for tracking product carriers. The application also incorporates a vendor verification status for each product. The system relies on three distinct use cases: Consumer, Carrier, and Producer, with varying activities for each role. Consumers place orders through either the mobile application or web platform.

The decentralized architecture spreads tasks among multiple nodes. As a result, it decreases the risk of execution failure at a single node. This strategy ensures no single node is solely responsible for the entire system, hence permitting a high level of reliability. Consumers have the ability to set an anticipated delivery time, ensuring flexibility and transparency in each delivery cycle. Once an order is placed, the system processes it, adds it to the request list, and notifies nearby carriers. This order request is visible to both carriers and vendors. If a carrier receives a request and can meet the expected delivery time, they accept the request, purchase the product from the vendor, and deliver it. Any registered user traveling from the product's origin to the order's destination can assume the role of a carrier.

Vendors authorize specific requests to guarantee product authenticity during the purchase process. Subsequently, carriers proceed to the consumer's location. Consumers can track carriers and verify product authenticity through the application. Upon delivery, carriers receive reviews and ratings from consumers. Completed order information is securely stored in the blockchain as encrypted hash data, preventing tampering or modification. The system's delivery fees depend on the distance between consumers and carriers, calculated by a backend algorithm. The following sections provide an in-depth explanation of the individual technologies and their functions.

**3.1  System Components**

The system and design have been precisely crafted to include blockchain technology and authentication, taking extensive study into account. The developed product delivery system consists of several precise technologies, including a private blockchain server that functions as a decentralized database for authenticated users, an application compatible with both mobile and web platforms for supply chain management, and a monitoring system that permits the tracking of the location of the product carrier. In addition, the application has a status for vendor verification for each product. There are three different use cases for the system: consumer, carrier, and producer. Each user participates in unique activities based on their assigned position.

A blockchain server is responsible for storing digital evidence and managing transactions; a smart contract system validates and generates digital contracts with hash data; a trusted directory service oversees certificates and facilitates their verification; a real-time server facilitates robust communication between the database and system; and a user-friendly application manages the system. In this part, the components of the framework are outlined, along with an explanation of digital evidence, data management certifications, and data management conditions.

### 3.1.1 Blockchain Server

In the aforementioned text, it was stated that there exist multiple data storage methods such as MySql, NoSql, SQLite, and others. After significant study and deliberation, it was determined that Blockchain is the appropriate solution for data gathering and storage because of its capacity to guarantee data security, manage decentralized databases, and give data transparency. Notably, a centralized database is highly dependent on the network connection and hence subject to failure.

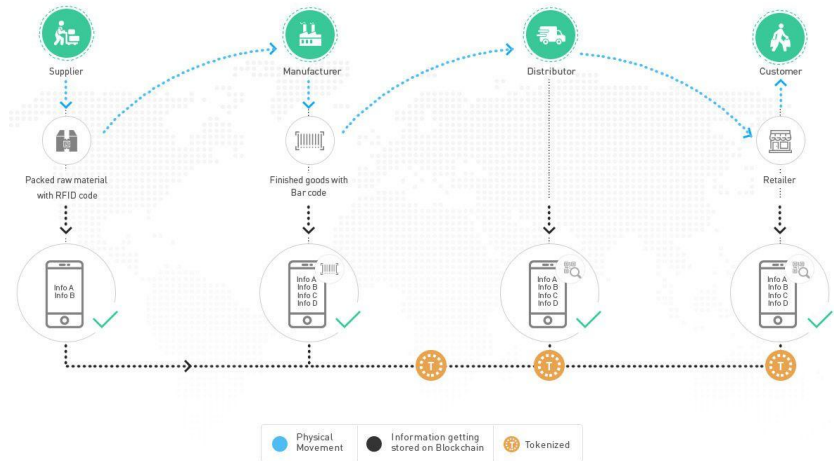

**Fig. 1.** Distributed Ledger Process

The use of a specialized blockchain server ensures the security and high availability of data via a decentralized database stored on a private network, notwithstanding the restricted potential of data backup. In addition, the validity of the product is assured since the product's provider authenticates it.

### 3.1.2 Smart Contract

The suggested method combines on-demand commodity delivery with Blockchain technology to improve transparency, security, and efficiency. Smart Contracts authenticate goods, track their whereabouts, and permit automatic payments, thereby minimizing the need for intermediaries and the likelihood of fraud. The immutable Blockchain record permits movement tracking, enhanced transparency, and theft prevention. Decentralization increases security and reduces the likelihood of hacking. Transactions that are expedited and cost-effective make distribution more accessible for small businesses and consumers.

Smart contracts organize information logically, allowing for automatic order classification and precise delivery. Upon delivery completion, they generate digital bills containing addresses, dates, and costs, which are recorded on the Blockchain, producing a transparent and unalterable transaction record

### 3.2 Procedure

Using decentralization, Proof-of-Stake consensus, cryptography, smart contract architecture, and immutable records, the system ensures data protection. These characteristics prevent hackers from altering data and validating transactions. Smart contracts prevent unauthorized access, hence decreasing fraud, errors, and criminal behavior. A permanent, tamper-proof record of transactions is preserved to ensure accountability and transparency. The combination of decentralization, Proof-of-Stake consensus, cryptographic methods, smart contract design, and record immutability strengthens data security overall. The full blockchain server procedure is as follows:

1. A new account will be established for the user in a particular node according to their function (Consumer, Carrier, or Producer) and they will be granted authorization to act as a sealer upon registration (for Consumers and Producers).
2. Upon receiving a new order, the carrier will send the transaction to Node2.
3. Once the product for a particular order has been delivered to the carrier, the authorized producer will finalize the transaction.
4. Upon delivery of the merchandise, the carrier will submit the transaction of delivery to the blockchain server.
5. The customer will finalize the transaction after getting the merchandise.

This is the sequence in which blocks will be added to the blockchain.

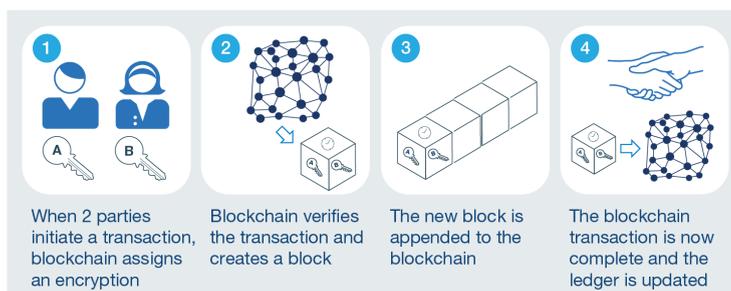

**Fig. 2.** Block Creation in Blockchain

### 3.3 Environment

The application is built using Flutter for Android and iOS, with a Node.js server API to interact with the application, Blockchain, and Firestore for storing data. The application facilitates communication between vendors, carriers, and consumers, with a focus on ensuring product authenticity throughout the product life cycle. Carrier tracking is done using GPS technology to monitor products and transportation, with the aim of reducing the risk of loss and enhancing security during transportation. The system relies on high-quality data, with a combination of real-time positioning and easy-to-use reporting applications to provide organizations with the confidence they need. NoSQL is used for storing general data in Firestore due to its low latency, large data volume, and flexible data models. A blockchain transaction is created for each server request.

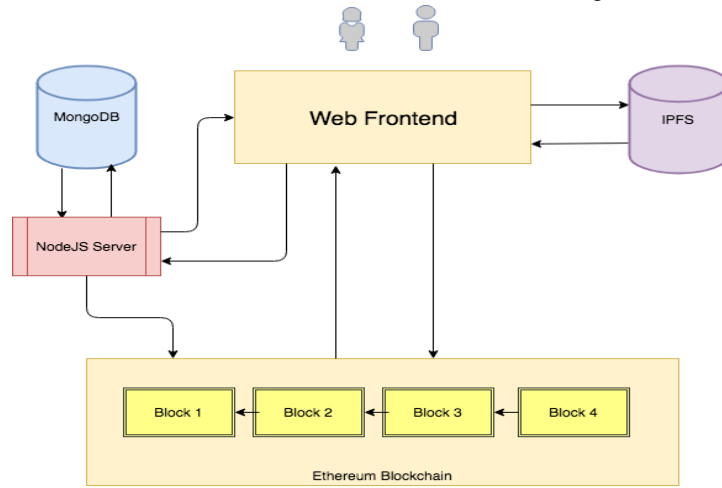

**Fig. 3.** Ethereum Transaction with Node.Js

## 4 Proposed System

The system has developed on different frameworks and languages for both websites and androids. Flutter is used for developing mobile applications and Laravel 7 framework for web applications. Flutter is based on dart programming language and Laravel 7 framework follows Model–view–controller (MVC) architecture. Frontend has developed on VueJS and Bootstrap whereas the backend has developed on Node.js that controls all data flow by building an API on Node.js. Node.js backend server is more stable and secure than others and is responsible for communicating with the fronted web, android app, BlockChain, and MongoDB to store data. NoSql has been used for storing the general data in Firestore for low latency, large data volume, and flexible data models.

Four user perspectives: users, guest regular users, registered carriers, and admin are available in the system. Users are registered regular users who can create an account in the system providing necessary information, can order products, can track their order via app or web page, view previous order history, edit their profile, can give a rating to the producers and carriers, and many more. A regular user can contact the carrier about the product and shipping status. Producers or sellers are guest users who can only see orders.

The website provides customers with features such as browsing and searching for products, adding products to their cart or wish list, and checking out by filling in billing details. Customers can also view their order history and track real-time order status through the user panel. Carriers can see available orders based on their location. The system has two types of administrators: admin and super admin. Both can control the system through the admin panel, which includes features such as adding coupons, updating the items list, and editing user information. Super admins have additional privileges such as adding and removing admin users. Admins can view order placement quantity, order status, and item availability list. They can also add, edit, or remove products from the system.

The system verifies users' information before approving transactions and sends a confirmation email to customers for each order placed. Carriers can only accept orders within a 10km radius and manage their delivery summary, profile, and order status. An admin is responsible for maintaining and managing the system. The system uses Restful API and stores sensitive information in a blockchain to prevent tampering. Ratings and reviews undergo cross-checking to ensure their authenticity. Orders have an expiry time, and shipping costs are automatically calculated based on distance and total units ordered.

A customer can access many features landing on the website's homepage such as products from the products page; however, customers can search for products manually. Individual product selection will redirect to specific pages. The system makes product information such as product price, rating, details, and location visible to consumers who can directly add products to their cart and update product quantity. Consumers can also keep products on their WishList. At the time of checkout, consumers need to fill up billing details along with the necessary information. Consumers can check order collection dates & duration, price, items, and relevant information after placing an order. Additionally, consumers can see previous order history. The user panel is operational on both the website and mobile apps with the advantage of real-time order tracking of product orders. Users can update their profile info, change passwords, etc. In the proposed system, carriers see all the available orders they can accept based on their nearest location.

The system has two different types: administrator and super admin. Admin panel has only access to the website. Both admins can control the entire system through this panel. The view of the system's dashboard, add coupons, update the items list, change system settings, and more features are available for the admin panel. Admins can edit user information from this page if necessary. Only the super admin has the advantage of adding new admin users, removing admin users, and many more. The admin can see order placement quantity, order status, and item availability list. The order status could be the shipment period, carriers' order received period and relevant information. Admin can also change order details as per requirement. Admin of the proposed system can add new items and edit existing product details. Besides, they can remove products from the system if items are not offered in-store.

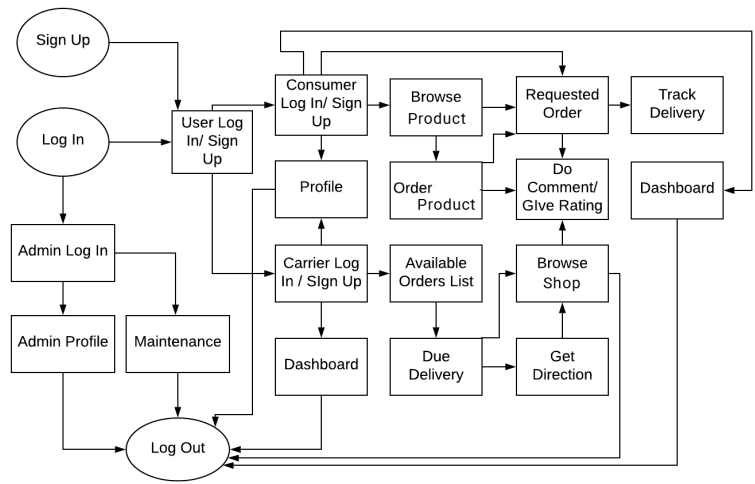

**Fig. 4.** Workflow Diagram

The transactions part of the system include initiation, validation via a consensus method, and activation of the smart contract to autonomously conduct the delivery procedure. Throughout the distribution process, a system coupled with the blockchain network is used for real-time tracking. At delivery completion, the smart contract automatically facilitates payment, and the invoice is updated within the blockchain, creating an immutable record of the transaction. This solution provides a secure and open way to transaction management.

The system will use an API to communicate with the server. The process of sending and receiving data will use Restful API. Blockchain technology is used to store sensitive information to prevent tampering with personal user data in the system. The rating and reviews of sellers and carriers would have to go through a cross-check facility before confirming to ensure whether they are legitimate or not. Every order has a customer-defined expiry time. As a result, the customer needs to reorder the product if no one accepts the order within the time period. In this situation, the shipping cost will be automatically generated based on the distance and total unit of the ordered product.

## 5 Design and Development of the Proposed System

This section details the design and development of our proposed system, beginning with its conception and ending with its actualization. Technique and technology employed guarantee a comprehensive understanding of the architecture and performance capabilities of the system.

### 5.1 Use-Case Diagram

The use-case diagram shows the interaction between users and the proposed system. It has three types of entities: Admin, Consumer, and Carrier. Log In and Log Out functionalities are compulsory for the system; however, other functionalities are distinct. The Carrier and Consumer users have to create their accounts whether the super admin only creates an admin account.

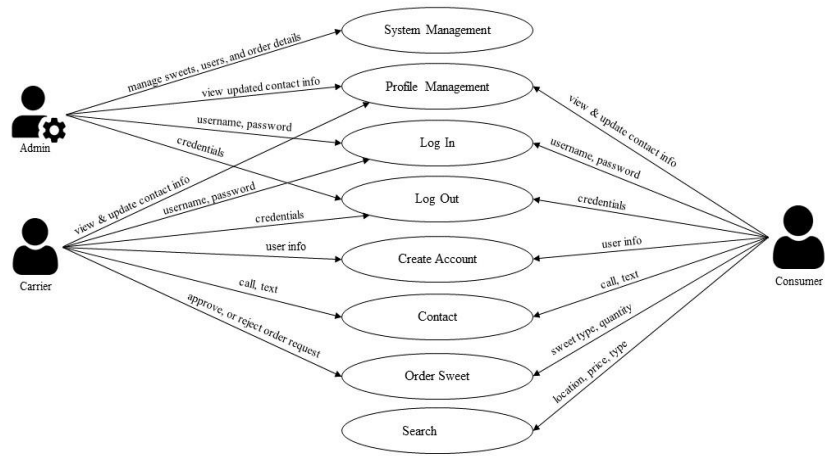

**Fig. 5.** Use Case Diagram

The admin is responsible for managing the whole system and can manage their profiles. Consumers can access all seven functionalities: Profile Management, Login, Log Out, Create Account, Order, Contract, and Search Items, and logging out of System Management. On the other hand, carriers can access the same functionalities except for the search for items.

### 5.2 User İnterface

The system has developed on different frameworks and languages for the mobile platform. Flutter has been used to build mobile, web, and desktop applications on a single codebase.

Node.js has been used for making the API that controls the whole system's data flow. The backend server of Node.js is responsible for communicating with the frontend web & android app. On the other hand, the runtime environment of Node.js helps to interact with the blockchain through applications. Cloud-based NoSQL document database 'Firestore' is used to store data for its low latency, large data volume, and flexible data models.

The developed system is crossed-platform and works smoothly on both mobile and web applications. The following images represent some User Interface designs of the mobile application of the proposed system.

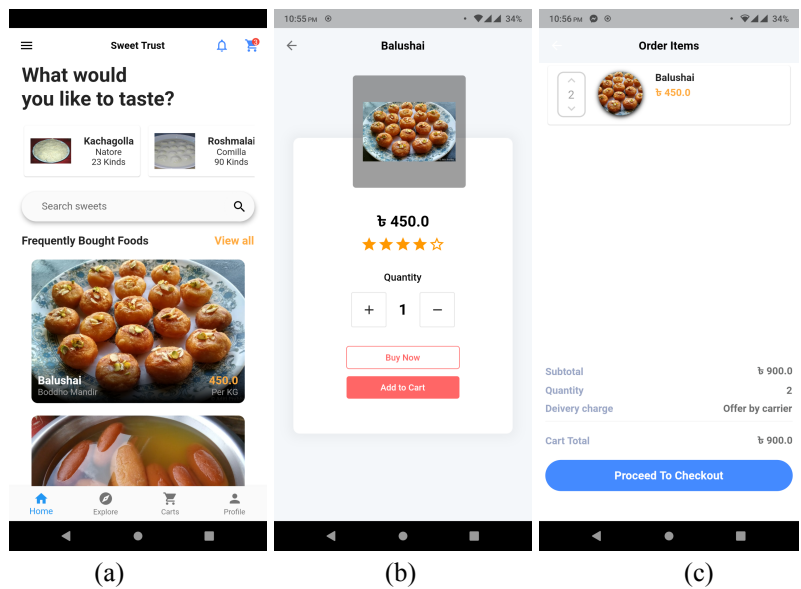

(a) (b) (c)

Fig. (a) is the Homepage of the customer account from where users can access many actions. Consumers can search and view all products from the product page. The search result will redirect them to the search result page, and an individual item selection will redirect to the selected product details page. The individual product item page contains the product image, price, rating, details, and location.

Consumers can either buy the product directly or can directly add the product's quantity to their cart Fig. (b). Consumers can save items in their cart for buying later. Additionally, they can update the item's quantity, and add products to save for later. The checkout page requires billing details and necessary information to complete the checkout Fig. (c).

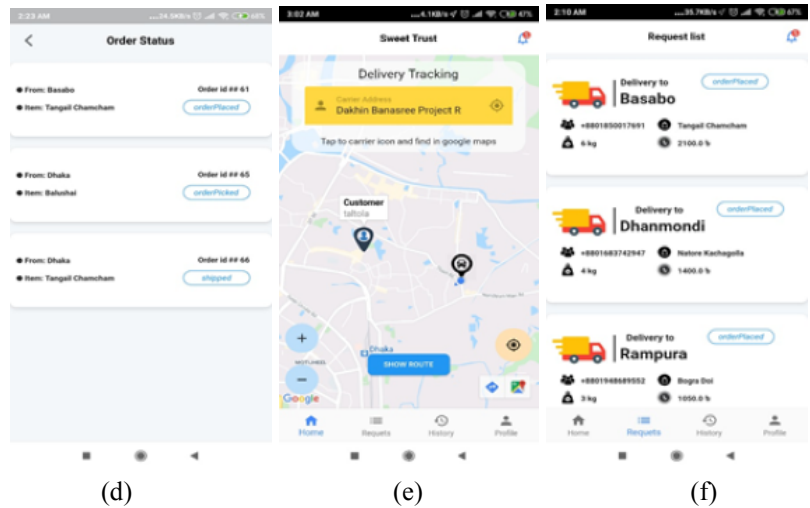

(d)  (e)  (f)

The order status page Fig. (d) is the previous order history of a consumer, such as pickup dates, prices, product items, etc. Real-time order tracking gives customers the advantage of tracking their order items in real-time through mobile apps Fig. (e). The Request List Fig. (f) page is only accessible from the carrier user end. Carriers can see all the available orders they can accept based on their location.

## 5  Performance Analysis

A customer can place multiple products at the same time from different places through the system. The system can give errors for the mentioned situation unless the exception could not be handled properly. As a result, the system divides a single order, counts every single item in a separate order considering the location of the product, and shows the order request to the carrier based on the product shop location. The current location of a carrier can not exit the radius of about 10.5 km to accept the order request. Hence, carriers must update their current location. The proposed system calls a Google Map Controller to reduce the issue of updating the location every 30 seconds. Besides that, multiple carriers might get the same delivery request based on their location. In this situation, the system's implemented process handles these types of conflict requests and gives those orders to the carrier based on the carrier rating and service overview record.

This product has some limitations with the features as this is a prototype. As a result, not all the features have been implemented yet except some of the basics. The system's analysis result shows that order tracking and status maintenance is a vital task. Data transfer from the app or Web App to the main server is a very crucial part of the system because phone numbers' improper handling can mess up the whole delivery chain and orders. The system uses a contact number that helps track orders to solve this circumstance. Interacting with the blockchain is crucial. A tiny data misconfiguration can shut down the blockchain server. Every data flow and authenticity of data origin is mandatory to re-check to avoid this situation.

## 6   Future Work & Conclusion

The study aims to build a transparent online platform to order and deliver an authentic product across Bangladesh using Blockchain, other security protocols, and algorithms. The reason for using Blockchain, security protocols, and algorithms is to ensure the transparency and security of the system. The current estimated food delivery market size is 10 million USD which could grow to over 14 million by the year 2025 [13]. The authentic product delivery market demand is high.

The system is now at the development level; however, the system's Blockchain server is going to be deployed in the public Ethereum network. The addition of some high-level algorithms and security protocols will make the system more robust and secure. A user must need to buy licenses to use the system. The Firestore storage plan will additionally help to develop the required settings and new attributes in future development. The reward option will help users and carriers to get a reward point based on their service. Users can use the reward point to get some discounts on products.

In the current food delivery system, it is hard to track whether customers are getting authentic products or they are not. However, the proposed solution gives the advantage of transparency and authenticity in the delivery process. The sooner stepping forward to the current food delivery market and providing genuine products would be the wisest. As a result, the whole system will bring significant change.

## References

1. Stadtler, H. (2015). Supply chain management: An overview. Supply chain management and advanced planning: Concepts, models, software, and case studies, 3-28.
2. Abbasi, W. A., Wang, Z., Alsakarneh, A. (2018). Overcoming SMEs financing and supply chain obstacles by introducing supply chain finance. HOLISTICA–Journal of Business and Public Administration, 9(1), 7-22.
3. Li, Z., Li, Z., Zhao, D., Wen, F., Jiang, J., Xu, D. (2017). Smartphone-based visualized microarray detection for multiplexed harmful substances in milk. Biosensors and Bioelectronics, 87, 874-880.
4. Yeo, S. F., Tan, C. L., Teo, S. L., Tan, K. H. (2021). The role of food apps servitization on repurchase intention: A study of FoodPanda. International Journal of Production

Economics, 234, 108063.
5. Ullah, G. W., Islam, A. (2017). A case study on Pathao: Technology-based solution to Dhaka's traffic congestion problem. Case studies in Business and Management, 4(2), 100-108.
6. Saad, A. T. (2021). Factors affecting online food delivery service in Bangladesh: an empirical study. British Food Journal, 123(2), 535-550.
7. Shipchain, https://docs.shipchain.io/docs/intro.html. Last accessed 4 Oct 2020
8. Hribernik, M., Zero, K., Kummer, S., Herold, D. M. (2020). City logistics: Towards a blockchain decision framework for collaborative parcel deliveries in micro-hubs. Transportation Research Interdisciplinary Perspectives, 8, 100274.
9. Badzar, A. (2016). Blockchain for securing sustainable transport contracts and supply chain transparency-An explorative study of blockchain technology in logistics
10. Toma, N. Z. (2021). An analysis on the use of operation and information technology management in a start-up logistic company.
11. Hwang, J., Lambert, C. U. (2008). The interaction of major resources and their influence on waiting times in a multi-stage restaurant. International Journal of Hospitality Management, 27(4), 541-551.
12. Muntasir, B. (2019). Meteoric rise of online food business. Dhaka Tribune.
13. Kader, R. (2020). The State of Online Food Delivery in Bangladesh at the Beginning of 2020: Subsidies Make True Demand Hard To Gauge, Future Start Up.)
14. Helo, P., Shamsuzzoha, A. H. M. (2020). Real-time supply chain—A blockchain architecture for project deliveries. Robotics and Computer-Integrated Manufacturing, 63, 101909.